\documentclass[journal]{IEEEtran}
\usepackage{algorithm,algpseudocode}
\usepackage{amsmath}
\usepackage{amssymb}
\usepackage{mathtools}
\usepackage{cite}
\def\BibTeX{{\rm B\kern-.05em{\sc i\kern-.025em b}\kern-.08em
    T\kern-.1667em\lower.7ex\hbox{E}\kern-.125emX}}

\usepackage{svg}
\usepackage{caption}
\usepackage{subcaption}
\usepackage{booktabs}

\usepackage{xcolor}
\begin{document}
%
\title{NEAT: \underline{N}on-lin\underline{e}arity \underline{A}ware \underline{T}raining for Accurate and Energy-Efficient Implementation of Neural Networks on 1T-1R Memristive Crossbars}

\author{Abhiroop Bhattacharjee$^{1*}$\thanks{\hspace{-3mm}$^*$ These authors have contributed equally \\ $\dagger$ This work was done while Lakshya was interning at Yale University}, Lakshya Bhatnagar$^{2*}$, Youngeun Kim$^{1*}$, and Priyadarshini Panda$^{1}$
\\$^{1}$Department of Electrical Engineering, Yale University, USA
\\
$^{2}$Department of Electrical Engineering, Indian Institute of Technology, Delhi, India$^{\dagger}$
\\
}

\maketitle

\begin{abstract}
%
%
Memristive crossbars have emerged as an energy-efficient component of deep learning hardware accelerators due to their compact and efficient Matrix Vector Multiplication (MVM) implementation. 
However, they suffer from non-idealities (such as, sneak paths) introduced by their circuit topology that degrades computational accuracy. 
A 1T-1R synapse, adding a transistor (1T) in series with the memristive synapse (1R), has been proposed to mitigate the non-idealities of crossbar.
We observe that the non-linear characteristics of the transistor affect the overall conductance of the 1T-1R cell which in turn affects the MVM operation. This 1T-1R non-ideality arising from the input voltage-dependent non-linearity is not only difficult to model or formulate, but also causes a drastic performance degradation of deep neural networks when mapped to such crossbars.  
In this paper, we analyse the non-linearity of the 1T-1R crossbar and propose a novel Non-linearity Aware Training (NEAT) method to address the non-idealities.
Specifically, we first identify the range of network weights, which can be mapped into the 1T-1R cell within the linear operating region of the transistor. 
After that, we regularize the weights of neural networks to exist within the linear operating range by using iterative training algorithm.
Our iterative training significantly recovers the classification accuracy drop caused by the non-linearity.
Moreover, we find that each layer has a different weight distribution and in turn requires different gate voltage of transistor to guarantee linear operation.
Based on this observation, we achieve  energy efficiency while preserving classification accuracy by applying heterogeneous gate voltage control to the 1T-1R cells across different layers.
Finally, we conduct various experiments on CIFAR10 and CIFAR100 benchmark datasets to demonstrate the effectiveness of our non-linearity aware training. Overall, NEAT yields $\sim20\%$ energy gain with less than $1\%$ accuracy loss (with homogeneous gate control) when mapping ResNet18 networks on 1T-1R crossbars.
\end{abstract}


%
\IEEEpeerreviewmaketitle

\section{Introduction}
%
%
%
%

\begin{figure}[t]
    \centering
    \includegraphics[width=0.8\linewidth]{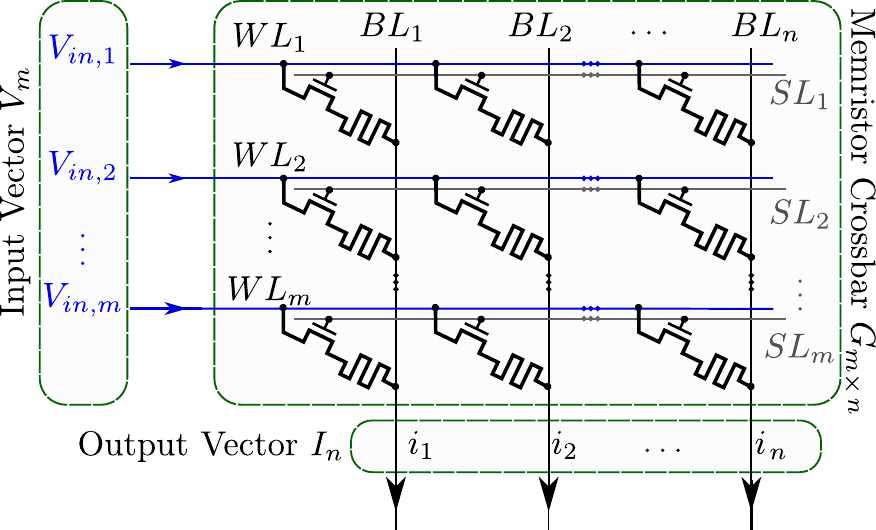}
    \caption{Illustration of 1T-1R Crossbar. A Transistor (T) with a NVM device (R)  at every junction of the Word-lines (WL) and bit-lines (BL). Select-lines (SL) are used to turn on transistors for selected rows. }
    \label{fig:1T1R}
     \vspace{-4mm}
\end{figure}

\IEEEPARstart{T}{he} last decade has seen the rise of Deep Neural Networks (DNNs) to solve many real-world problems. Their promising real-world application and growing resource requirements have lead to researchers focusing on dedicated hardware accelerators. As the CMOS digital hardware advancement cannot keep up with the growing computational needs of DNNs \cite{xu2018scaling}, Non-Volatile-Memory (NVM) based crossbars have emerged as a compact and efficient realization for performing the Matrix-Vector-Multiplication (MVM) operations of DNNs in the analog domain \cite{ankit2019puma,jain2020rxnn}. Fig.~\ref{fig:1T1R} illustrates an $m\times n$ crossbar. Here, a $1 \times m$ input vector of voltages ($V_i$) interacts with a matrix of NVM conductances ($G_{ij}$) to produce the output current $I_j = \Sigma_{i=1}^{m} V_i*G_{ij}$. Hence, the currents from the $n$ columns of the crossbar constitute the output vector of the MVM operation. Especially, 1T-1R NVM crossbars (Fig.~\ref{fig:1T1R}) have been widely studied since the transistor in series with the NVM device can help mitigate sneak paths and the incorrect programming of the NVM device induced by noise~\cite{li2017sneak, wang2014ferroelectric}.

However, the presence of a transistor in the synapse introduces various non-idealities pertaining to the non-linear I-V characteristics of a transistor. These non-idealities are data-dependent~\cite{chakraborty2020geniex} and DNNs when mapped onto such crossbars suffer computational accuracy losses. Most of the previous works~\cite{jain2020rxnn, lee2020learning, chen2015mitigating, liu2014reduction, liu2015vortex} have proposed strategies and frameworks to model and mitigate data-independent non-idealities (primarily resistive non-idealities and NVM device variations) pertaining to 1R crossbar arrays to improve on the accuracy of the mapped DNNs. However, none of these works have proposed methods to mitigate the transistor-induced non-linearities (data-dependent) in 1T-1R crossbars for the weight-to-conductance mapping in an energy-efficient manner. Recent work \textit{GenieX}~\cite{chakraborty2020geniex} provides a neural network based framework to model both data-dependent and data-independent non-idealities for a crossbar array of 1T-1R synapses. However, it lacks transferability to crossbars with different specifications and requires re-training of the neural network to model the non-idealities. Thus, approaches towards ensuring efficient mapping of DNNs onto 1T-1R crossbars in an energy-constrained environment has not been well explored. This is highly crucial because 1T-1R crossbars are increasingly becoming prospective candidates for deployment in extremely resource-constrained environment such as IoT devices, drones among others.



In this work, we provide a new perspective on the energy-efficient implementation of DNNs on 1T-1R crossbars. Specifically, we focus on the gate-voltage of the transistor, on which the power consumption of a 1T-1R crossbar system is found to largely depend for a given set of analog input voltages. We observe that 1T-1R synapse has approximately linear characteristics when sufficiently high gate-voltage is supplied. However, if the gate-voltage is low (resource-constrained scenario for low power operation), then the synapse starts exhibiting non-linearity. This non-linearity becomes even more pre-dominant when synapses are programmed to higher conductance values. This non-linear characteristic of 1T-1R synapse degrades the accuracy of DNNs when mapped onto crossbars. 

To address this problem, we propose a Non-linearity Aware Training (NEAT) technique.
Given the gate-voltage and trained network weights, we first compute the range of conductances for linear operation of the 1T-1R cell via SPICE simulations.
Based on this range, we force the weight parameters to be within the linear regime.
To this end, we train the networks by iterative training consisting of two steps:
(1) approximating the trained weight parameters in the non-linear regime to the boundary of the linear regime; (2) re-training the network with modified weights.
We repeat these steps so that a greater number of weight parameters can lie in the linear regime.
Also, in this work, we propose two gate-voltage settings in order to implement an energy-efficient crossbar.
We can set the same gate-voltage across all layers (homogeneous) or different gate-voltage for each layer (heterogeneous) of the DNN.
The homogeneous approach results in high energy gains but can incur significant accuracy losses.
The heterogeneous setting addresses this problem by searching layer-wise gate-voltage that guarantees a small accuracy drop while achieving energy-efficiency.
Both the settings can be applied in NEAT and we show their efficiency through extensive experiments.




In summary, we focus on mitigating the non-linear and data-dependent non-idealities introduced on addition of a transistor to the crossbar-synapses by operating with optimal selection of transistor gate-voltage. We take a device-agnostic approach assuming that the NVM device can be programmed to a given conductance. Once the non-linear effects are countered, one can use the existing methods to counter other data-independent non-idealities, such as interconnect parasitics, NVM device variations, \textit{etc.}~\cite{jain2020rxnn, lee2020learning, chakraborty2020geniex}. Thus, our proposed NEAT is complementary to prior works dealing with mitigating or modelling non-idealities in analog crossbars.

The key contributions of this work are as follows:

\begin{itemize}
    \item Comprehensive analysis through SPICE simulations to determine factors affecting crossbar power and transistor-induced non-linearities in 1T-1R synapse
    \item Determination of the maximum permissible value of effective synaptic conductance for linear operation for a given transistor gate-voltage. In essense, our work unleashes gate-voltage as a control knob to limit the non-idealities encountered in a 1T-1R crossbar array while yielding energy-efficiency. 
    \item Propose Non-linearity Aware Training (NEAT) for accurate and energy-efficient implementation of DNNs on 1T-1R memristive crossbar.
    By using NEAT, we achieve  $\sim$20\% energy gain on ResNet18 architecture on CIFAR10, CIFAR100 datasets while having a tolerable accuracy loss ($\sim$1\%).
\end{itemize}

\section{Background and Preliminary Work}

To understand the effects of introducing the access transistor (or selector) in the synapse, we performed extensive SPICE simulations using the 1T-1R configuration with different input voltage, conductance and gate-voltage ranges. The selector devices are based on PTM 45nm CMOS technology model. We have considered a memristive device with $R_{ON}=30k\Omega$ and $R_{OFF}=300k\Omega$. For a given supply voltage $V_{supply}$, the input to any word-line ($V_{in}$) will be in the range $0\le V_{in}\le V_{supply}$.

\subsection{1T-1R Power Analysis}
\begin{figure}
    \centering
    \includegraphics[width=0.5\linewidth]{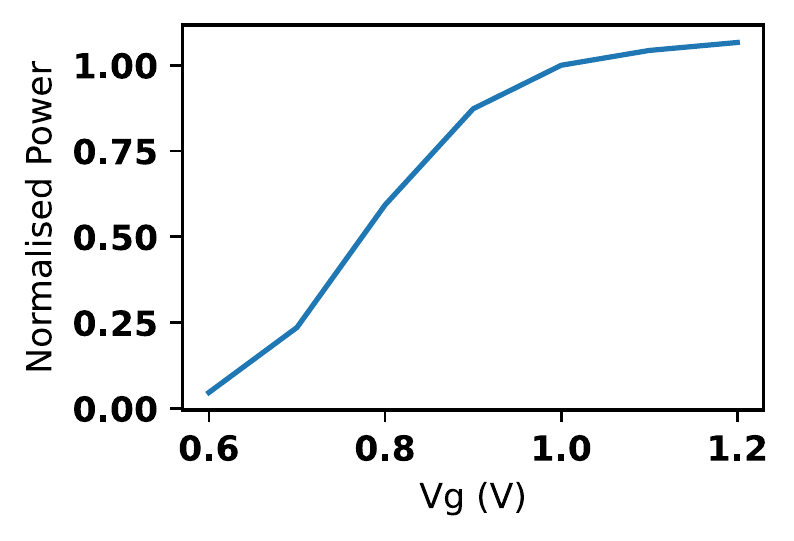}
    \caption{Average 1T-1R synapse Power for Monte Carlo simulation. Normalised for $V_g = 1V.$
    }
    \label{fig:1t1rPower}
     \vspace{-4mm}
\end{figure}

Before diving into non-linearity considerations, we characterise the role of transistor gate-voltage $V_g$ in the per synapse power consumption. We performed a Monte Carlo simulation on a $8\times{8}$ 1T-1R crossbar with weights drawn drawn from a normal distribution and mapped to conductance,  
and inputs drawn from a uniform distribution. Average power per synapse is shown in Fig \ref{fig:1t1rPower}.
The results suggest that $V_g$ plays a considerable role in determining power consumption.
\subsection{Analysis of Transistor Induced Non-Linearity}
For a crossbar in the 1R configuration, the weights $W$ of the DNN are directly mapped to a memristor conductance state ($G_M = 1/R_M$). On the other hand, in the 1T-1R configuration, $W$ is mapped to the effective conductance $G_{eff} = 1/(R_M + R_t)$, where $R_t$ is the equivalent resistance due to the transistor. 
The non-linearities in the 1T-1R crossbars arise due to the dependence of $R_t$ on $V_{in}$
. Note, $V_{in}$ is proportional to the neuronal activation values of the DNN which varies with the input. Hence, these are referred to as data-dependent non-idealities.

\begin{figure}
    \centering
    \includegraphics[width=\linewidth]{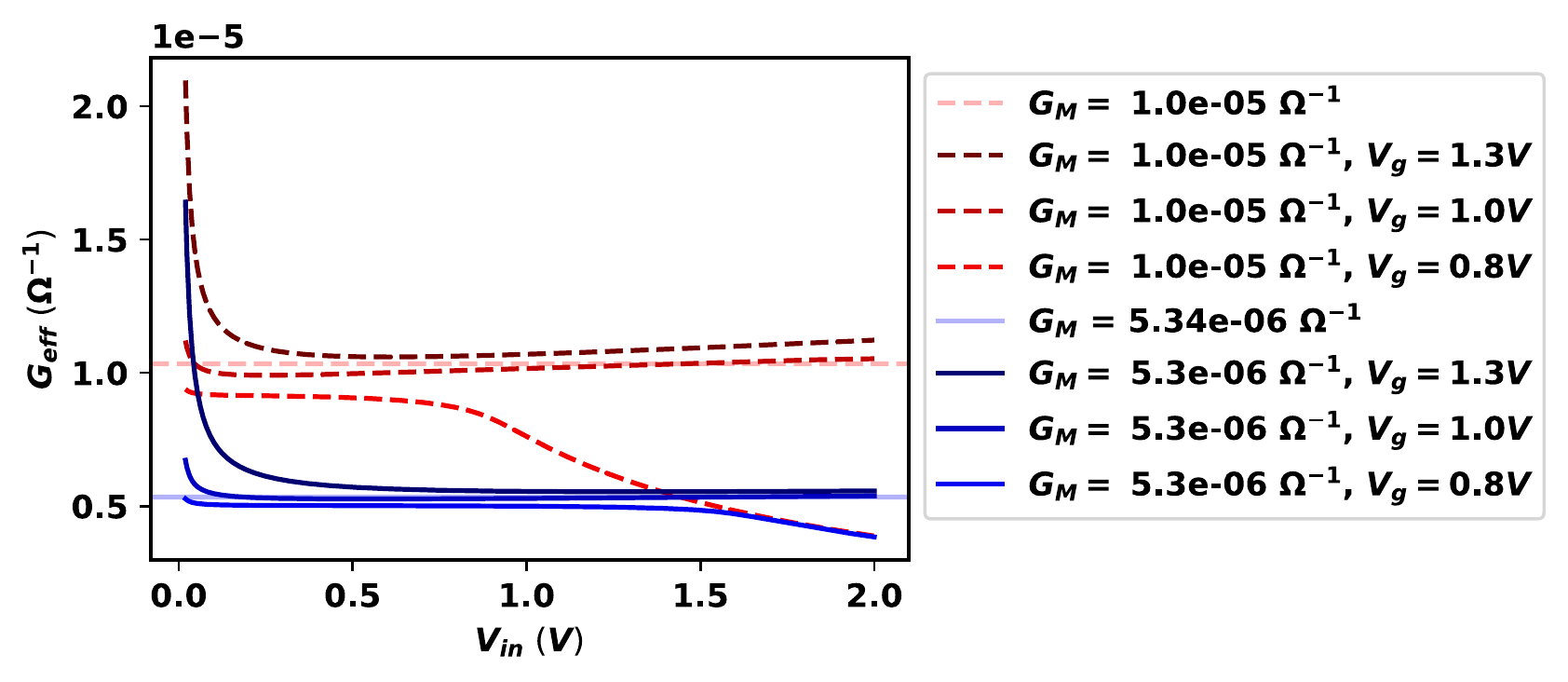}
    \caption{Characterising $G_{eff}$ for 1T-1R synapse. Variation in $G_{eff}$ with $G_M$,$V_{in}$, $V_g$ is shown. 
    }
    \label{fig:geff}
     \vspace{-4mm}
\end{figure}

As shown in Fig \ref{fig:geff}, the effective conductance $G_{eff}$ is a function of NVM conductance $G_M$, input voltage $V_{in}$, and gate-voltage $V_g$, i.e. 
\begin{equation}
G_{eff}=f_1(G_M,V_{in},V_g) \label{eqn:geff:actual}
\end{equation}

The key takeaway from Fig. \ref{fig:geff} is that $V_{in}$ has significant influence on $G_{eff}$ for both smaller and larger $G_M$ values.
For lower $V_{in}$
, the $G_{eff}$ non-ideality can be attributed to the leakage current in the transistor. Thus, in low $V_{in}$ ranges, lower $V_g$ operation of transistor will nullify the leakage current. Hence, even though $G_{eff} \to \infty$ when $V_{in} \to 0$, the distortion in $G_{eff}$ is negligible for low $V_g$. On the other hand, leakage current is not negligible for higher $V_g$, hence the distortions in $G_{eff}$ cannot be neglected. Further, at higher values of $V_{in}$ and low $V_g$ operation, the transistor begins to shift its region of operation from linear to saturation, as a result of which we find deviations in the value of $G_{eff}$. At the juncture of the linear and saturation regions of operation, $R_t$ increases and thus, there is a larger voltage drop across it causing a dip in the value of $G_{eff}$. This can be seen in Fig. \ref{fig:geff} for $V_{in}> 1$ when $V_g = 0.8$. Note, for high $V_g \ge 1$, such stauration effects occur  at a much higher $V_{in} > 2 $ thus making it irrelevant.

\subsection{Finding 1T-1R Linear Regime}
\label{subsec:LinearBehaviour}
In the ideal scenario, we would expect $G_{eff}$ to be a horizontal line parallel to the x-axis for any given value of $G_M$ in case of Fig \ref{fig:geff}. Since we keep selector gate-voltage constant for an MVM operation, we want: 
\begin{equation}
G_{eff}=f_2(G_M,V_g) \label{eqn:geff:ideal}
\end{equation}
To omit data-dependence, our objective here is to find out the range of parameters for which Eqn. \ref{eqn:geff:ideal} is a reasonable approximation for Eqn. \ref{eqn:geff:actual}. That is, there should be a weak or no dependence of $G_{eff}$ on $V_{in}$. Once we get rid of this data-dependence, we can infer that controlling $V_g$ will allow us to operate the transistor in a linear regime where, $G_M\approx k* G_{eff}$, where $k$ is a scalar. 

\begin{figure}
    \centering
    \begin{subfigure}{0.9\linewidth}
        \includegraphics[width=0.45\linewidth]{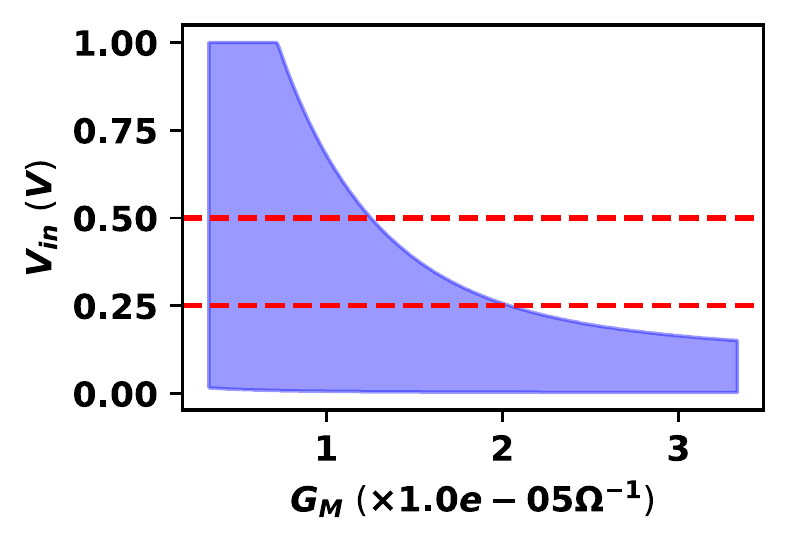}
        \hfill
        \includegraphics[width=0.4203\linewidth]{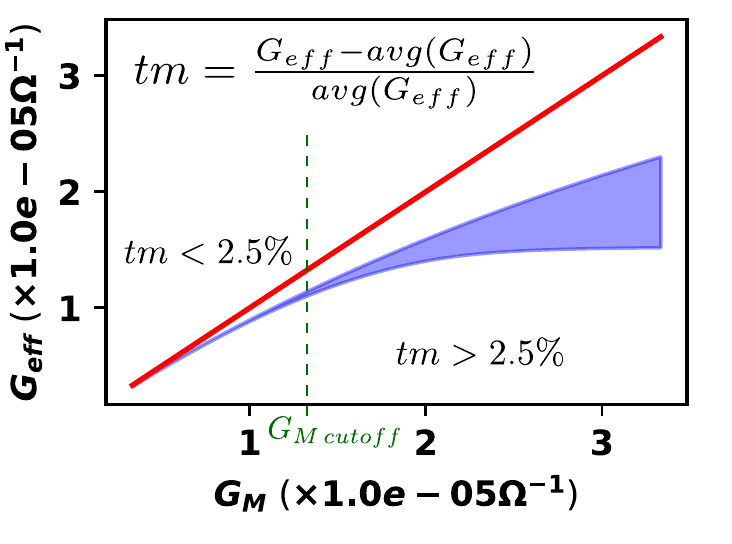}
        \caption{$V_g=0.8V$}
        \label{fig:Vbound:0.8}
    \end{subfigure}
    \begin{subfigure}{0.9\linewidth}
        \includegraphics[width=0.45\linewidth]{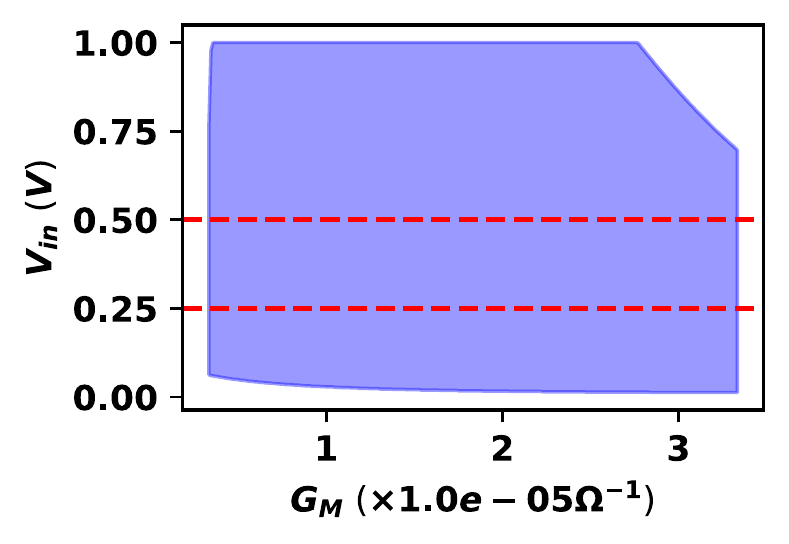}
        \hfill
        \includegraphics[width=0.4203\linewidth]{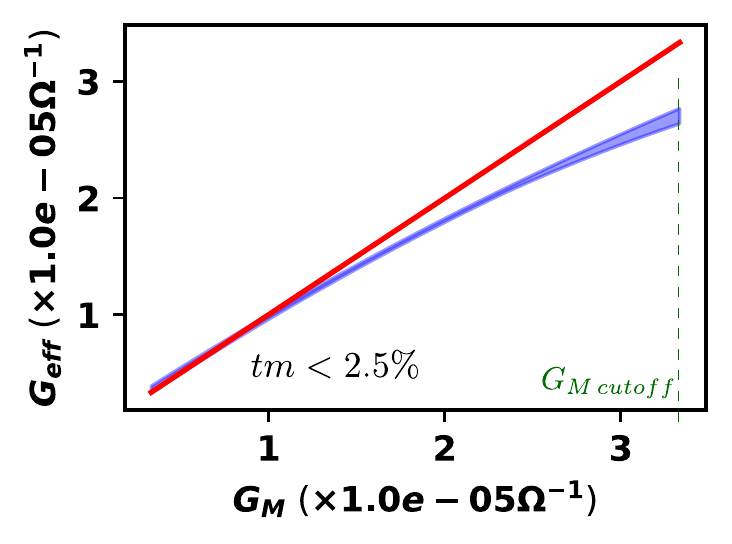}
        \caption{$V_g=1.0V$}
        \label{fig:Vbound:1.0}
    \end{subfigure}
    \begin{subfigure}{0.9\linewidth}
        \includegraphics[width=0.45\linewidth]{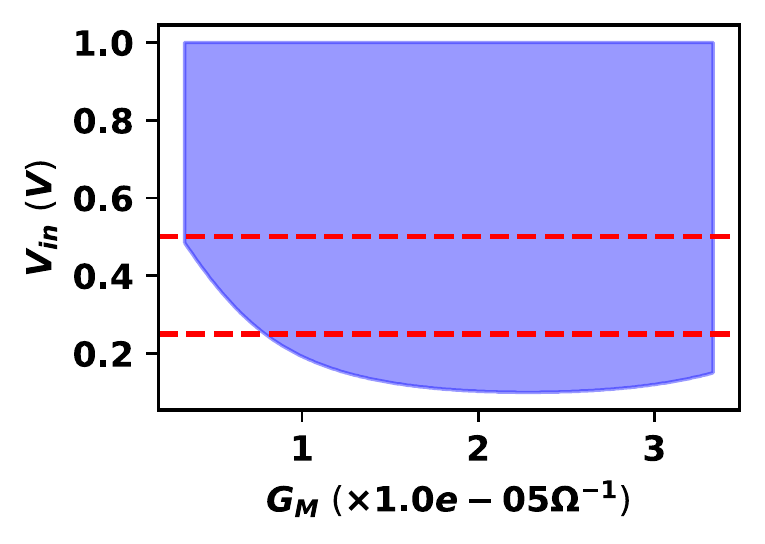}
        \hfill
        \includegraphics[width=0.4203\linewidth]{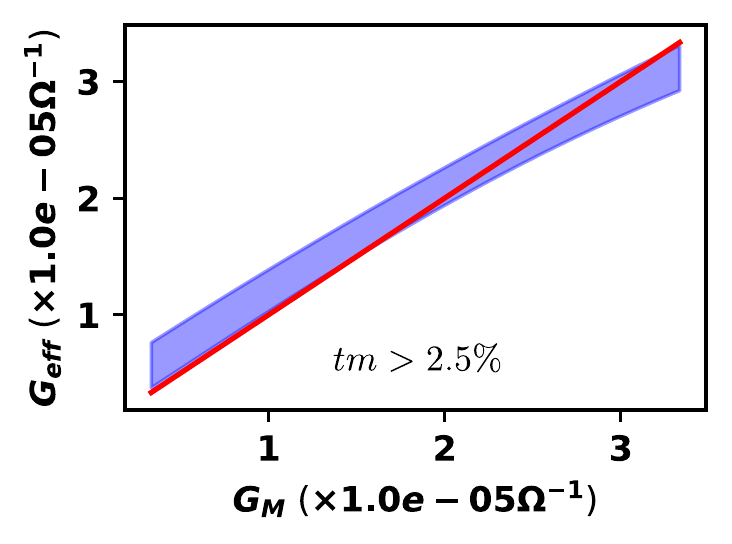}
        \caption{$V_g=1.3V$}
        \label{fig:Vbound:1.3}
    \end{subfigure}
    \caption{(Left) $G_M$ vs range of $V_{in}$ for which linearity can be assumed. The blue shaded region essentially marks the region where linearity is observed, $G_{eff} \approx k*G_M$. $V_{supply} / V_{in} =0.25,\, 0.5 V$ is marked in the plots for convenience. (Right) $G_M$ vs range of $G_{eff}$ for $V_{supply}=0.5 V$ is shown. Ideal $G_{eff} = G_M$ line (red) drawn for convenience. The blue shaded line shows the overall deviation of $G_{eff}$ from $G_M$ for an input voltage range $0<V_{in}<0.5$. $tm$ is the tolerance metric. The $G_{M\, cutoff}$ values for $tm <2.5\%$ are annotated. For $V_g =1.3$, $tm > 2.5\%$ for the entire range of $G_M$ values.
    }
    \label{fig:Vbound}
     \vspace{-4mm}
\end{figure}

Fig. \ref{fig:Vbound} (Right) illustrates the 1T-1R non-linearity where we show the $G_{eff}$ values obtained for a range of $G_M$ for $0<V_{in}<0.5$ across different $V_g$. For a particular $G_M$ (especially larger values) we observe the spread of $G_{eff}$ becomes more prominent. Higher the spread of the blue region, higher is the data-dependence of $G_{eff}$ on $V_{in}$ for the given $G_M$. So, restricting the spread in $G_{eff}$ will curb non-linearity of the 1T-1R synapse. This will also ensure that the transistor operates in the linear regime with a constant $R_t$ (that is data-independent). Thus, we define a tolerance metric (\textit{tm}) to quantify the spread or deviation in $G_{eff}$ as shown in Fig. \ref{fig:Vbound} (Right). 
Fig. \ref{fig:Vbound} (Left) further illustrates the range of $V_{in}$ for which linearity can be assumed for a given $G_M$ across different $V_g$ values.

At low $V_g = 0.8V$, unavailability of higher $V_{in}$ values for higher values of $G_M$ in Fig. \ref{fig:Vbound:0.8} (Left) can be attributed to transistor saturation. These results support the observations in prior work \textit{GenieX} \cite{chakraborty2020geniex}. The authors found higher supply voltage ($V_{supply}$) of $0.5V$ (or higher $V_{in}$) yields higher non-ideality than lower supply voltage $V_{supply}=0.25V$. As seen from Fig. \ref{fig:Vbound:0.8}, we can explain this as an artefact of the 1T-1R non-linearity.
At high $V_g = 1.3V$, unavailability of lower $V_{in}$ for lower $G_M$ in Fig \ref{fig:Vbound:1.3} (Left) can be attributed to high leakage current. Both of these effects are not noticed in Fig \ref{fig:Vbound:1.0} (Left).
To make sure that NEAT is capable of handling the worst case, we use $V_{supply}=0.5V$ in all our experiments.

The above discussion indicates that there is an upper limit to $V_g$ for achieving desirable linear characteristics for a given $V_{in}$ range.
\subsection{Defining Conductance cut-off ($G_{eff\, cutoff}$)}
\label{subsec:GeffCutoff}
\begin{figure}
    \centering
    \includegraphics[width=0.8\linewidth]{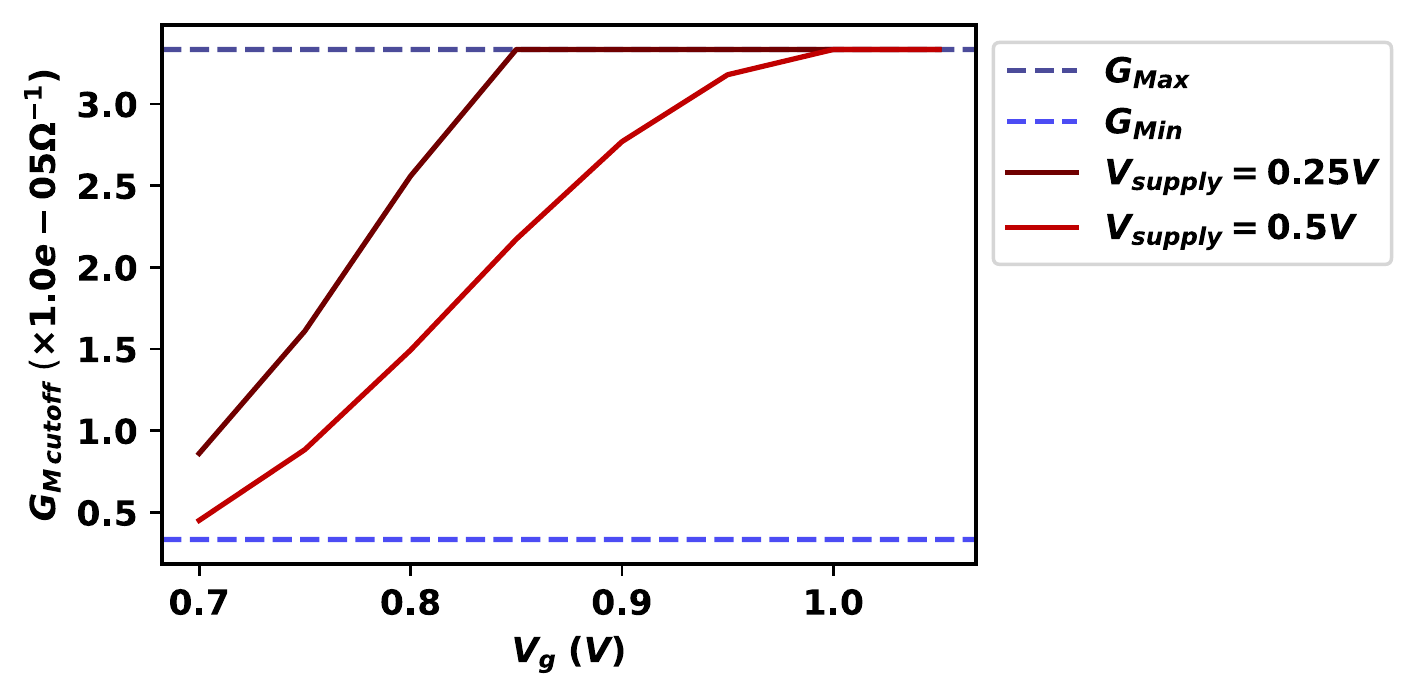}
    \caption{The change of $G_{M\, cutoff}$ with respect to $V_g$.}
    \label{fig:GeffCutoff}
     \vspace{-4mm}
\end{figure}

Based on our above discussion, we observe that for a given set of $V_g$, $V_{supply}$ and tolerance metric, an upper bound cut-off value ($G_{eff\, cutoff}$) exists for which the 1T-1R synapse exhibits linear characteristics.
We term the corresponding NVM device state as $G_{M\, cutoff}$. $G_{M\, cutoff}$ has been annotated in Fig. \ref{fig:Vbound:0.8} for a tolerance value $<2.5\%$. For $V_g =0.8V$ and $1V$, $G_{M\, cutoff}$ values are around $1.25 \times 10^{-5}\Omega^{-1}$ and $3.34 \times 10^{-5}\Omega^{-1}$ respectively. For $V_g =1.3V$, we don't have any cutoff for the given tolerance metric. This implies that we cannot operate the 1T-1R synapse at $V_g=1.3V$. Thus, from this analysis, we obtain the $G_{eff\, cutoff}$ values that can be used to determine the corresponding range of software DNN weights which will ensure linear operation of 1T-1R synapse after mapping. Fig \ref{fig:GeffCutoff} shows the $G_{M\, cutoff}$ vs. $V_g$ plot for $2.5\%$ tolerance metric and $V_{supply}=0.25V,~ 0.5V$.


Note, the focus of the above analysis is on the correctness of the Multiply-and-Accumulate (MAC) operation that happens at the synapse level based on $V_g$, $V_{supply}$ and $G_{eff\, cutoff}$ values. While extending this analysis to a crossbar, feasibility of all these parameters should be checked. For e.g. there would be a limit to leakage current permissible for each bit-line dictated by the sensing device, which would in-turn set a per synapse limit on leakage current, essentially adding a constraint on $V_g$. Thus, at an array level, new $V_g$, $V_{supply}$ and $G_{eff\, cutoff}$ constraints need to be calculated. However, the overall methodology of obtaining the cut-off parameters for linear operation will be the same as discussed above.

\section{Non-linearity Aware Training (NEAT)}

\begin{figure}[t]
    \centering
    \includegraphics[width=0.85\linewidth]{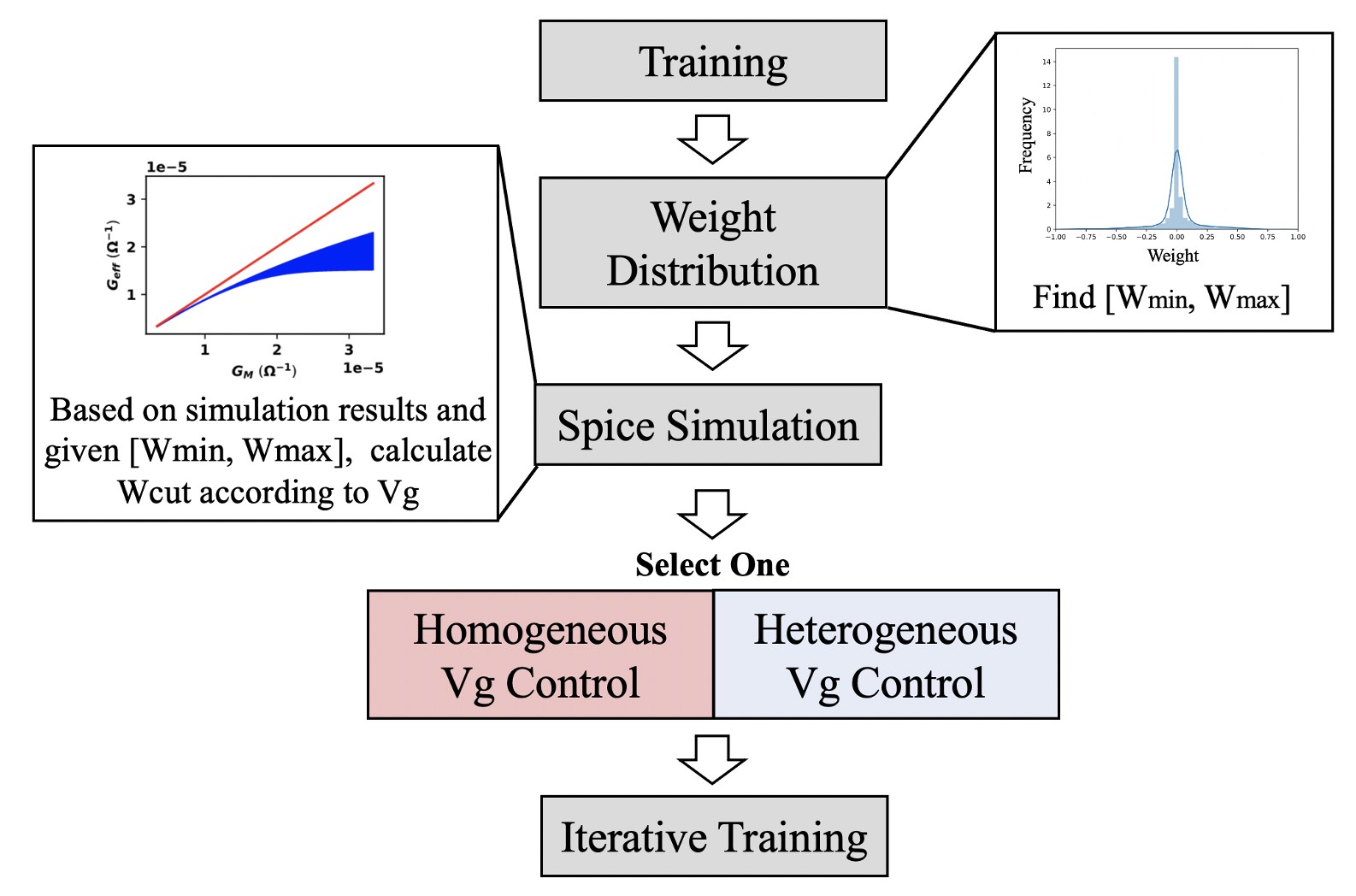}
    \caption{ Overall flow of NEAT.
}
    \label{fig:method}
     \vspace{-4mm}
\end{figure}

The non-linearity of $G_{eff}$ can induce performance degradation in DNNs when the corresponding trained weights $W$ are mapped onto crossbars. To mitigate performance losses, we convert $G_{eff\, cutoff}$ (as determined in Section~\ref{subsec:GeffCutoff}) to obtain the corresponding $W_{cut}$ for the software DNN. Then, we restrict all the weights ($W$) of the DNN in the interval $[-W_{cut}, W_{cut}]$ as shown in (\ref{eq:Wcut}):

\begin{equation}
    W_{map} =
\begin{cases}
 W    & |W| \leq W_{cut}  \\
    W_{cut}  & W > W_{cut}  \\ 
     -W_{cut} & W < -W_{cut}.
\end{cases}
\label{eq:Wcut}
\end{equation}
From Eqn. \ref{eq:Wcut}, we observe that for the linear regime ($|W| \leq W_{cut}$ which corresponds to $G_{eff} \approx G_M$), the software weight parameters can be mapped linearly onto the crossbars.
While, for the non-linear regime ($|W| > W_{cut}$ that corresponds to deviation of $G_{eff}$ from $G_M$), $W$ is clipped at $W_{cut}$.
The objective of NEAT is to restrict the weight parameters to be within the linear regime for the given gate-voltage $V_g$ of the transistor.
Fig. \ref{fig:method} illustrates the overall flow of the  NEAT process.

\begin{figure}[t]
    \centering
    \includegraphics[width=0.85\linewidth]{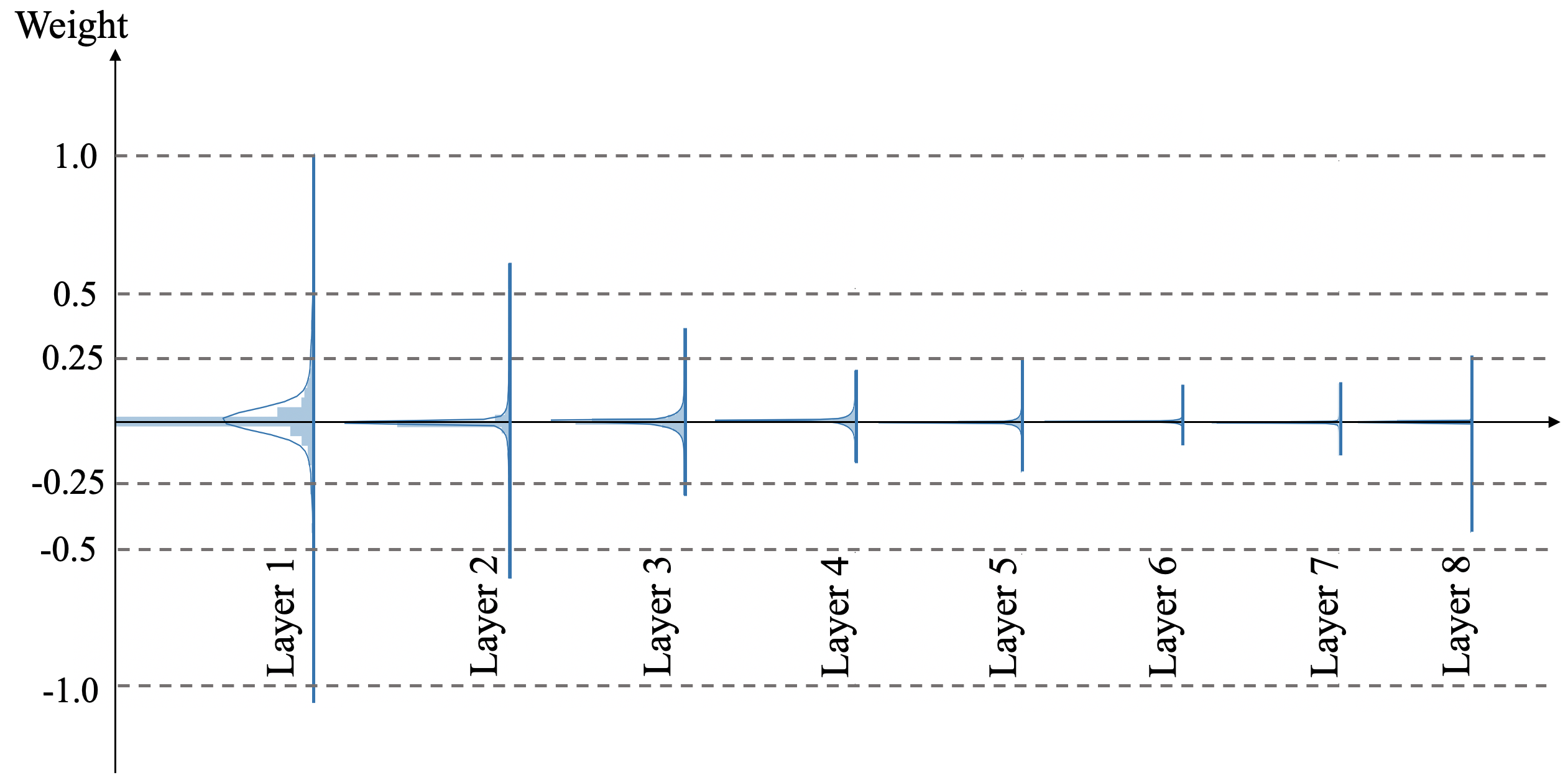}
    \caption{Illustration of layer-wise weight distribution. We use a VGG11 architecture on CIFAR10.}
    \label{fig:weightdistribution}
     \vspace{-4mm}
\end{figure}

\subsection{Training DNNs and Obtaining $W_{cut}$}


We train the network with cross-entropy loss on the classification dataset. 
From the trained networks, we can find the maximum and the minimum weight values, namely  $W_{min}$ and $W_{max}$.
After that, given all the weight parameters,
we calculate the $W_{cut}$ value for a given $V_g$ value by determining $G_{eff\, cutoff}$ using SPICE simulations as described in Section \ref{subsec:GeffCutoff}. We set the search range of $V_g$ as $[0.7, 1.0]$ with interval of 0.05.

\subsection{$V_g$ Control Schemes: Homogeneous and Heterogeneous}

With the above-stated relationship between $W_{cut}$ and $V_g$,  we  suggest  two  different strategies for $V_g$ selection.

\textbf{Homogeneous $V_g$ control:}
We use the same $V_g$ value for $W_{cut}$ computation across all layers of the DNN.
This method ensures high energy-gains but can incur significant accuracy losses. It is hard to predict how much accuracy degradation will be incurred
without accessing the validation or test dataset.
In other words, for a given $W_{cut}$ (related to a given $V_g$),
we can identify the percentage of weights that will operate in the non-linear 1T-1R regime after mapping. But we cannot identify the accuracy drop without accessing the test dataset. 
This might be crucial for the applications where accuracy should be preserved.

\textbf{Heterogeneous $V_g$ control:}
To address the above-mentioned problem, we suggest a heterogeneous approach where each layer has a different $V_g$ allocated.
We observe that each layer of the DNN has a different weight distribution, as shown in Fig. \ref{fig:weightdistribution}.
Using this observation, we allocate a low value of $V_g$ (corresponding to a smaller $G_{eff\, cutoff}$ and hence, $W_{cut}$) for the layers having a smaller range weights (e.g., layer 5 $\sim$ layer 7), otherwise, we allocate a high value of $V_g$ (e.g., layer 1).
By doing this, we can obtain energy-efficiency by guaranteeing that most of the weights lie in the linear regime. Note, we set the $V_g$ and $W_{cut}$ for all layers based on the weight distribution without having access to the test or validation data. 


Algorithm 1 shows the strategy of selecting the optimal $V_g$ for the heterogeneous approach.
The purpose is to set the $V_g$ configuration to extract significant energy-efficiency. Any accuracy loss incurred in the DNN with setting the $V_g$ and corresponding $W_{cut}$ is recovered by iterative training.
To this end,  we first define the maximum absolute value of weights in layer $l$ (lines 1-3).
Then, we search the optimal $V_g$ value in ascending order.
Based on the given $V_g$, we calculate $W_{cut}$. 
If the current $W_{cut}$ covers all the weights values, we set the last $V_{g\_prev}$ as the optimal $V_g$ at layer $l$ and stop searching (lines 4-12).
It is worth mentioning that setting $V_{g\_prev}$ or $W_{cut}$  distorts the weight distributions that causes accuracy decline. This loss can be minimized by iterative training described in the next subsection. We further note that heterogenous $V_g$ selection does not require any data. 




\begin{algorithm}[t]\small
    \caption{Heterogeneous $V_g$ Searching}
  \textbf{Input}: DNN weights ($W$); $V_g$ search list ($S =$ [0.7, 1.0; 0.05])\\
  \textbf{Output}:  layer-wise gate voltage $V_{g}$
  \begin{algorithmic}[1]
    \For{$l \gets 1$ to $L$}
        \State{\% Find the maximum weight representation at layer $l$}
        \State{$l.W_{r} = max(|l.W_{min}|, |l.W_{max}|)$}
        \For{$V_g$ in $S$} 
            \State{\% Define weight cutoff based on  $V_{g}$ }
            \State{$l.W_{cut} \leftarrow (V_g, l.W)$}
            \State{\% Find the optimal $V_{g}$ }
            \If{$l.W_{cut} > l.W_{r}$  } 
                \State{$l.V_{g} \leftarrow V_{g\_prev}$}
                \State{\textbf{break}}
            \EndIf
            \State{$V_{g\_prev} \leftarrow V_g$}
        \EndFor
    \EndFor
  \end{algorithmic}
    
\end{algorithm}

\subsection{Iterative Training}

In NEAT, after setting the optimal $V_g$ and $W_{cutoff}$ values from homogeneous or heterogeneous gate control, we then transform the weights of the DNN. If we use lower values of $V_g$ which do not cover all weight ranges, the weight distribution gets altered, resulting in accuracy degradation.
To address this issue, we propose iterative training which consists of two steps, as shown in Algorithm 2. Step 1 is essentially restricting the weights of the DNN ($W$) in the suitable cut-off regime as per Eqn. (\ref{eq:Wcut}). In Step 2, we retrain the networks iteratively for a couple of epochs to recover any accuracy loss incurred from Step 1.
We repeat these two steps so that greater number of weights in the network can be located in the linear regime when mapped onto crossbars. 
By carrying out extensive experiments, we show that the iterative training significantly improves the performance in lower $V_g$ scenarios.

\begin{algorithm}[t]\small
    \caption{Iterative Training}
  \textbf{Input}: The number of iteration ($N$), DNN weights ($W$);  layer-wise gate voltage $V_{g}$  \\
  \textbf{Output}: Trained DNN weights ($W$)
  \begin{algorithmic}[1]
    \For{$n \gets 1$ to $N$}
        \State{\% STEP 1: Apply $W_{cut}$ to $W$ according to $V_{g}$}
        \State{$W_{cut} \leftarrow V_{g}$}
        \State{$W \leftarrow (W, W_{cut})$}
        \State{\% STEP 2: Train $W$}
        \State{$W \leftarrow train(W)$}
    \EndFor
  \end{algorithmic}
   
\end{algorithm}

\section{Experimental Results}

We conduct our experiments on PyTorch with VGG11 \cite{simonyan2014very} and ResNet18 \cite{he2016deep} architectures on CIFAR10 and CIFAR100 datasets \cite{krizhevsky2009learning}.
For retraining process (Step 2 in Algorithm 2), 
we use Adam optimizer with learning rate $10^{-5}$.
For all experiments, the number of iteration ($N$ in Algorithm 2) is 30 which implies low overhead for retraining.

\subsection{Analysis on Homogeneous $V_g$ Selection}

\begin{figure}[t]
    \centering
    \includegraphics[width=0.83\linewidth]{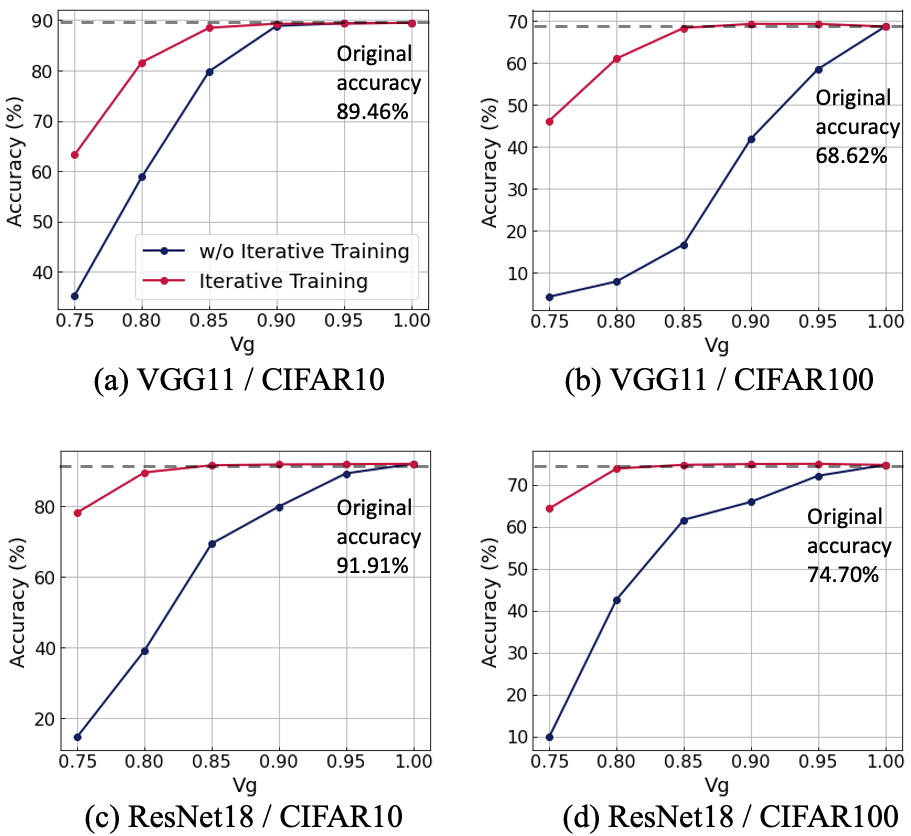}
    \caption{Classification accuracy with respect to $V_g$. }
    \label{fig:homo_acc}
\end{figure}

\begin{figure}[t]
    \centering
    \includegraphics[width=0.83\linewidth]{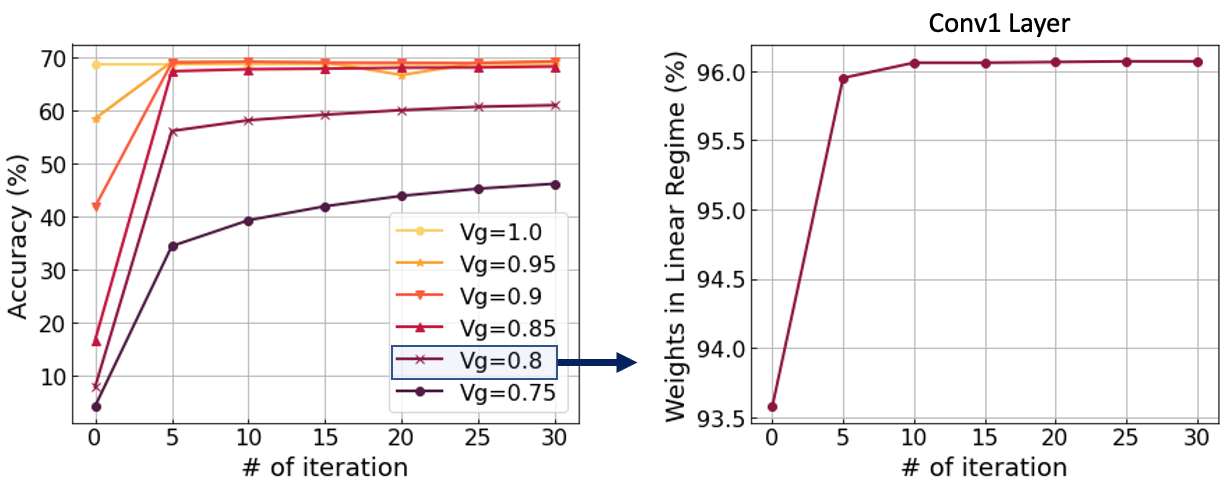}
    \caption{Classification accuracy with respect to the number of iterative training. We use VGG11 on CIFAR100.}
    \label{fig:num of iteration}
\end{figure}

In Fig. \ref{fig:homo_acc}, we change $V_g$ from 0.75 to 1.0 and report the classification accuracy.
The results show that low  $V_g$ induces low $W_{cut}$ and in turn decreases performance when DNN weights are restricted to $W_{cut}$ regime. However, using iterative training recovers the performance degradation. 
Especially, for a ResNet18 architecture, using iterative training shows improvement over 50\%  in terms of accuracy at $V_g = 0.75$.
Moreover, with iterative training, VGG11 and ResNet18 networks almost maintain their classification accuracy in the range of $V_g = [0.85, 1.0]$ and $V_g = [0.8, 1.0]$, respectively.
%
To further validate the effectiveness of iterative training, we provide the accuracy with respect to the number of iterations.
Fig. \ref{fig:num of iteration} shows that the classification accuracy improves as the number of iterations ($N$ in Algorithm 2) increases.
This is because iterative training forces the weights to be in the linear regime of operation.
To validate this, we plot the percentage of weights in the linear regime at the first convolution layer ($V_g  = 0.8$ case).
The results demonstrate that the majority of weights ($\sim$ 96\%) are located in the linear regime after 30 iterations in case of iterative training.
Other layers also show similar results.

\subsection{Analysis on Heterogeneous  $V_g$ Selection}

\begin{figure}[t]
    \centering
    \includegraphics[width=1.0\linewidth]{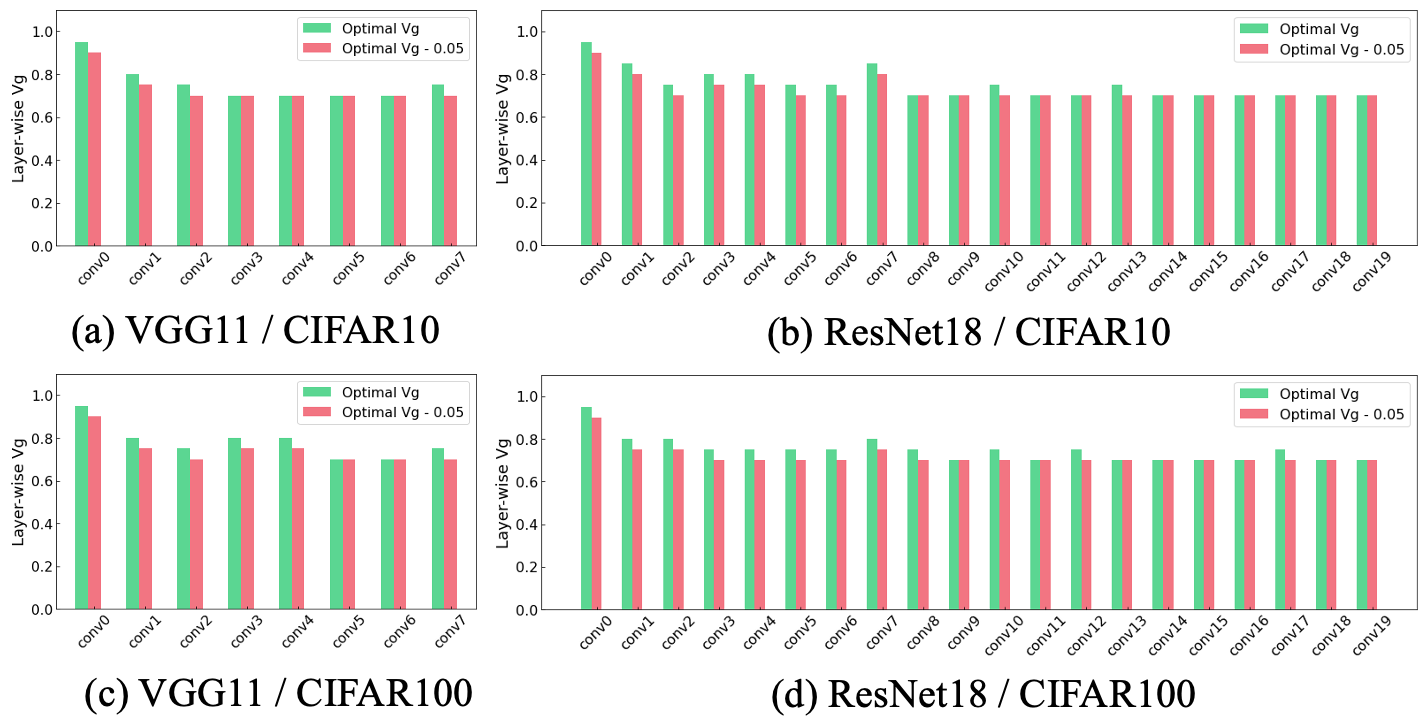}
    \caption{$V_g$ configurations of heterogeneous setting. ``Optimal $V_g$" denotes the $V_g$ configuration from searching algorithm (Algorithm 1).
    Also, in ``Optimal $V_g - 0.05$", we reduce $V_g$  of all layers by $0.05$.
    }
    \label{fig:hetero_vg}
\end{figure}

Based on Algorithm 1, we obtain the heterogeneous layerwise $V_g$ configuration, and define this as “Optimal $V_g$”. 
In order to study the energy-accuracy trade-off, we conduct experiments on the more energy-efficient configuration, named “Optimal $V_g - 0.05$”. Here, we take the individual layerwise $V_g$ values obtained from Algorithm 1 and further subtract 0.05V as shown in Fig. \ref{fig:hetero_vg}.
%
In Fig. \ref{fig:hetero_vg}, we observe that high $V_g$ values are required for the input layer, whereas low $V_g$ values are required for the intermediate layers.
Table \ref{table: hetero_performance} presents the classification accuracy in case of the heterogeneous approach. Just plainly restricting the DNN weights in the cut-off range based on “Optimal $V_g$” configuration yields $<0.3\%$ accuracy drop, even without any iterative training across all models and datasets.
For “Optimal $V_g-0.05$” (Fig. \ref{fig:hetero_vg}), drastic performance degradation is observed since most of the $V_g$ values are set to the minimum value 0.7 which lowers the $W_{cut}$. Iterative training in this case increases the accuracy by $2-5\%$. It is worth mentioning that we can find a more fine-grained solution by setting the search interval (in this case, $0.05$) to a smaller value.  


\begin{table}[t]
\addtolength{\tabcolsep}{0pt}
\centering
\caption{Classification accuracy of heterogeneous setting.}
\vspace{-1mm}
\resizebox{1\linewidth}{!}
{
\begin{tabular}{lcccccccc}
\toprule
  Model & Dataset & Iterative Training & Optimal $V_g$  & Optimal $V_g - 0.05$  \\
\midrule
  VGG11 & CIFAR10 &  No & 88.05  & 78.55 \\
  VGG11 & CIFAR10 &  Yes  & 88.74 & 83.51  \\
\midrule
  Res18 & CIFAR10 &   No &90.60  & 55.67  \\
  Res18 & CIFAR10 &  Yes  & 90.58 & 59.23  \\
\midrule
  VGG11 & CIFAR100 &   No & 68.42  & 63.64  \\
  VGG11 & CIFAR100 &  Yes  & 68.88 & 66.06 \\
 \midrule
  Res18 & CIFAR100 &   No & 72.99  & 27.62  \\
  Res18 & CIFAR100 &  Yes  & 73.60 & 29.22 \\
 \bottomrule
\end{tabular}%
}
\label{table: hetero_performance}
\vspace{-4mm}
\end{table}

\begin{figure}[t]
    \centering
    \includegraphics[width=0.8\linewidth]{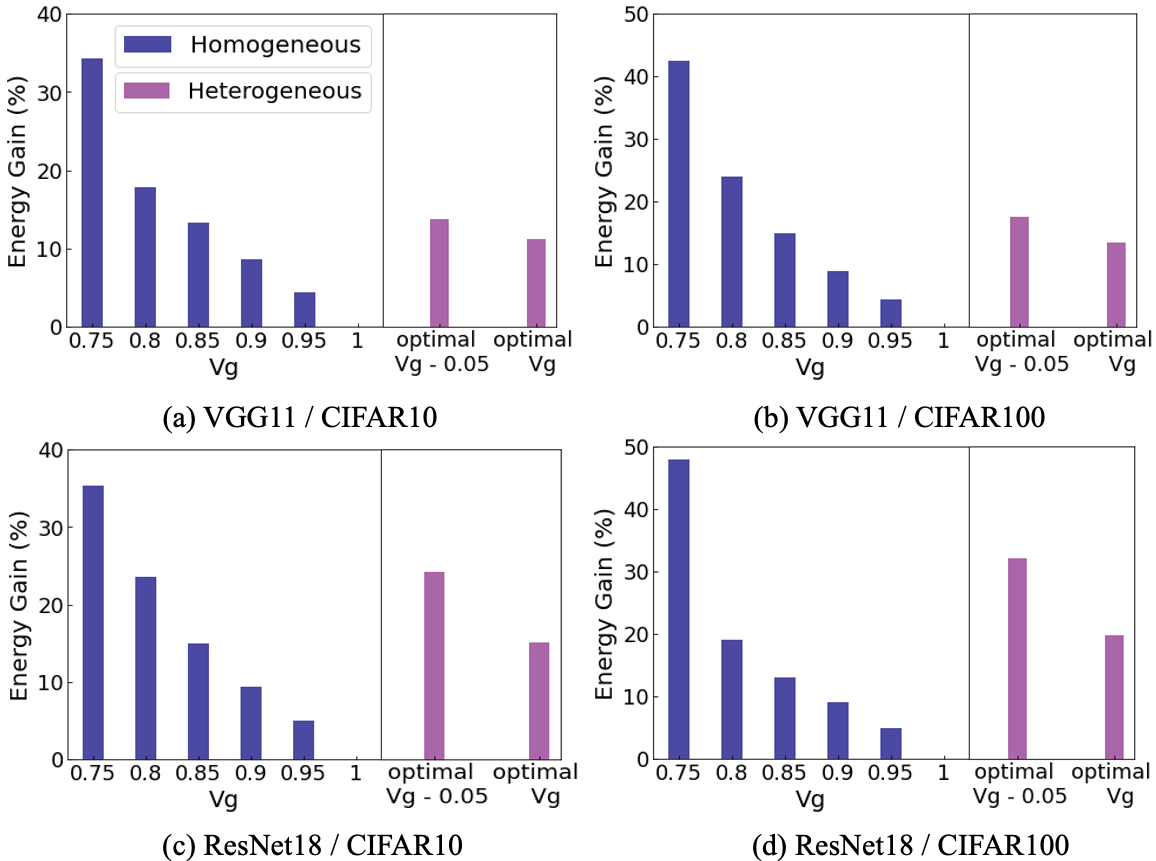}
    \caption{Energy gain from homogeneous and heterogeneous approaches. }
    \label{fig:energy_comparison}
\end{figure}

\subsection{Analysis on Energy Efficiency}

Finally, we present the energy efficiency of various configurations in Fig. \ref{fig:energy_comparison}.
We measure the energy consumption in 1T-1R crossbars following previous work~\cite{ankit2017resparc}. We use the energy computed for homogeneous gate-control scenario for $V_g = 1.0$ as baseline against which energy gain (\%) are shown.
For the homogeneous setting, we
can achieve high energy gain by simply reducing $V_g$.
Especially, we can achieve $\sim$ 23\% energy gain at $V_g = 0.8$ on ResNet18 architecture with CIFAR10 while suffering minimal accuracy loss ($\sim$1.5\%).
However, selecting a very low value for $V_g$ such as $V_g=0.75$ induces huge performance degradation (Fig. \ref{fig:homo_acc}).
For the heterogeneous setting, we obtain over 10\% energy gain for all experiments.

\section{Conclusion}
We propose a novel training method that takes into account the non-linear characteristics of the transistor (selector) in 1T-1R crossbars.
We analyse this non-linearity in low transistor gate-voltage scenario using both algorithm and hardware perspectives.
Moreover, we mitigate the effect of non-linearity by iterative training.
Our experimental  results  demonstrate that our proposed NEAT technique achieves energy-efficiency while preserving the classification accuracy of the DNNs.
Our work can impact the future deployment of 1T-1R crossbar in extremely resource-constrained environment.


\section*{Acknowledgement}
This work was supported in part by the National Science Foundation (Grant\#1947826), and the Amazon Research Award.

\bibliographystyle{IEEEtran}
\bibliography{reference}

\begin{thebibliography}{10}
\providecommand{\url}[1]{#1}
\csname url@samestyle\endcsname
\providecommand{\newblock}{\relax}
\providecommand{\bibinfo}[2]{#2}
\providecommand{\BIBentrySTDinterwordspacing}{\spaceskip=0pt\relax}
\providecommand{\BIBentryALTinterwordstretchfactor}{4}
\providecommand{\BIBentryALTinterwordspacing}{\spaceskip=\fontdimen2\font plus
\BIBentryALTinterwordstretchfactor\fontdimen3\font minus
  \fontdimen4\font\relax}
\providecommand{\BIBforeignlanguage}[2]{{%
\expandafter\ifx\csname l@#1\endcsname\relax
\typeout{** WARNING: IEEEtran.bst: No hyphenation pattern has been}%
\typeout{** loaded for the language `#1'. Using the pattern for}%
\typeout{** the default language instead.}%
\else
\language=\csname l@#1\endcsname
\fi
#2}}
\providecommand{\BIBdecl}{\relax}
\BIBdecl

\bibitem{xu2018scaling}
X.~Xu, Y.~Ding, S.~X. Hu, M.~Niemier, J.~Cong, Y.~Hu, and Y.~Shi, ``Scaling for
  edge inference of deep neural networks,'' \emph{Nature Electronics}, vol.~1,
  no.~4, pp. 216--222, 2018.

\bibitem{ankit2019puma}
A.~Ankit, I.~E. Hajj, S.~R. Chalamalasetti, G.~Ndu, M.~Foltin, R.~S. Williams,
  P.~Faraboschi, W.-m.~W. Hwu, J.~P. Strachan, K.~Roy \emph{et~al.}, ``Puma: A
  programmable ultra-efficient memristor-based accelerator for machine learning
  inference,'' in \emph{Proceedings of the Twenty-Fourth International
  Conference on Architectural Support for Programming Languages and Operating
  Systems}, 2019, pp. 715--731.

\bibitem{jain2020rxnn}
S.~Jain, A.~Sengupta, K.~Roy, and A.~Raghunathan, ``Rxnn: A framework for
  evaluating deep neural networks on resistive crossbars,'' \emph{IEEE
  Transactions on Computer-Aided Design of Integrated Circuits and Systems},
  2020.

\bibitem{li2017sneak}
T.~Li, X.~Bi, N.~Jing, X.~Liang, and L.~Jiang, ``Sneak-path based test and
  diagnosis for 1r rram crossbar using voltage bias technique,'' in
  \emph{Proceedings of the 54th Annual Design Automation Conference 2017},
  2017, pp. 1--6.

\bibitem{wang2014ferroelectric}
Z.~Wang, W.~Zhao, W.~Kang, Y.~Zhang, J.-O. Klein, and C.~Chappert,
  ``Ferroelectric tunnel memristor-based neuromorphic network with 1t1r
  crossbar architecture,'' in \emph{2014 International Joint Conference on
  Neural Networks (IJCNN)}.\hskip 1em plus 0.5em minus 0.4em\relax IEEE, 2014,
  pp. 29--34.

\bibitem{chakraborty2020geniex}
I.~Chakraborty, M.~F. Ali, D.~E. Kim, A.~Ankit, and K.~Roy, ``Geniex: A
  generalized approach to emulating non-ideality in memristive xbars using
  neural networks,'' \emph{arXiv preprint arXiv:2003.06902}, 2020.

\bibitem{lee2020learning}
S.~Lee, G.~Jung, M.~E. Fouda, J.~Lee, A.~Eltawil, and F.~Kurdahi, ``Learning to
  predict ir drop with effective training for reram-based neural network
  hardware,'' in \emph{2020 57th ACM/IEEE Design Automation Conference
  (DAC)}.\hskip 1em plus 0.5em minus 0.4em\relax IEEE, 2020, pp. 1--6.

\bibitem{chen2015mitigating}
P.-Y. Chen, B.~Lin, I.-T. Wang, T.-H. Hou, J.~Ye, S.~Vrudhula, J.-s. Seo,
  Y.~Cao, and S.~Yu, ``Mitigating effects of non-ideal synaptic device
  characteristics for on-chip learning,'' in \emph{2015 IEEE/ACM International
  Conference on Computer-Aided Design (ICCAD)}.\hskip 1em plus 0.5em minus
  0.4em\relax IEEE, 2015, pp. 194--199.

\bibitem{liu2014reduction}
B.~Liu, H.~Li, Y.~Chen, X.~Li, T.~Huang, Q.~Wu, and M.~Barnell, ``Reduction and
  ir-drop compensations techniques for reliable neuromorphic computing
  systems,'' in \emph{2014 IEEE/ACM International Conference on Computer-Aided
  Design (ICCAD)}.\hskip 1em plus 0.5em minus 0.4em\relax IEEE, 2014, pp.
  63--70.

\bibitem{liu2015vortex}
B.~Liu, H.~Li, Y.~Chen, X.~Li, Q.~Wu, and T.~Huang, ``Vortex: variation-aware
  training for memristor x-bar,'' in \emph{Proceedings of the 52nd Annual
  Design Automation Conference}, 2015, pp. 1--6.

\bibitem{simonyan2014very}
Simonyan \emph{et~al.}, ``Very deep convolutional networks for large-scale
  image recognition,'' \emph{arXiv:1409.1556}, 2014.

\bibitem{he2016deep}
He \emph{et~al.}, ``Deep residual learning for image recognition,'' in
  \emph{IEEE CVPR, 2016}, 2016.

\bibitem{krizhevsky2009learning}
A.~Krizhevsky, G.~Hinton \emph{et~al.}, ``Learning multiple layers of features
  from tiny images,'' 2009.

\bibitem{ankit2017resparc}
Ankit \emph{et~al.}, ``Resparc: A reconfigurable and energy-efficient
  architecture with memristive crossbars for deep spiking neural networks,''
  \emph{arXiv:1702.06064}, 2017.

\end{thebibliography}



%








\end{document}